\documentstyle[aps,twocolumn,tighten,epsfig]{revtex}
\begin{document}

\draft
\title{ DENSITY OF BLOCH WAVES AFTER A QUENCH }

\author{Jacek Dziarmaga\thanks{E-mail: {\tt ufjacekd@thrisc.if.uj.edu.pl}}}
\address{ Institute of Physics, Jagiellonian University,
          Reymonta 4, 30-059 Krak\'ow, Poland}
\date{September 10, 1998}
\maketitle
\tighten

\begin{abstract}
{\bf Production of Bloch waves during a rapid quench is studied by analytical 
and numerical methods. The density of Bloch waves decays exponentially with 
the quench time. It also strongly depends on temperature. Very few textures are 
produced for temperatures lower than a characteristic temperature proportional 
to the square of the magnetic field. }
\end{abstract}
\vspace*{0.5cm}

   Topological defects appear in many condensed matter systems
\cite{davydov}. They are also believed to play a role in cosmology
\cite{vilenkin} and in nuclear physics \cite{skyrme}. Kibble \cite{kibble}
pointed out that the defects can be produced in significant numbers during
a rapid second order transition from the disordered to the symmetry
broken phase. The speed of light was the key factor in his estimate
of the defect density just after a quench. Zurek \cite{zurek} stressed
the importance of the nonequilibrium Ginzburg-Landau dynamics of the
order parameter. According to the Kibble-Zurek scenario the density
of kinks, vortices or monopoles scales like a certain defect-dependent
power of the quench rate. The theory was verified in several
experiments \cite{condmat}.

   Both theory and experiment concentrated so far on hedgehogs,
which have a nontrivial asymptote at spatial infinity. Textures are
a wide class of topological defects with a constant asymptote.
A texture which arises in a $D+1$ component spin system has $D$ spatial
dimensions. As the spin vector is constrained to have definite magnitude,
it lays on a $D$ dimensional sphere. An external magnetic field prefers, 
say, downward spin orientation. A spin field has finite energy
if its spins point down at spatial infinity. The $D$ dimensional
space is compactified to a $D$ dimensional sphere. A finite energy spin
field is a map $S^{D}\rightarrow S^{D}$. If the map has a nonzero winding
number, then it can not be continuously deformed to a uniform spin-down
ground state. 

   The simplest example of a topological texture is a Bloch wave
in one spatial dimension. Its spin is an in-plane arrow
of definite length, which points down far from the wave. 
As we pass through the (anti-)texture from the left to the right the spin 
makes the full (anti-)clockwise rotation. The Bloch wave has a definite size
as a result of balance between the Zeeman and the exchange energy.

   In this paper we study the production of Bloch waves during a rapid quench. 
We believe that our results are qualitatively correct for all
topological textures like skyrmions in planar ferromagnetic systems
(e.g. quantum Hall ferromagnets \cite{QHE}), magnetic bubbles or Hopfions
in three dimensional ferromagnets \cite{bubbles} and
baryons in the Skyrme model \cite{skyrme}. In addition, Bloch waves are
formally equivalent to fluxons in long Josephson junctions \cite{ljj}.

    The Bloch wave is topologically stable because its spin is constrained to 
lay on a circle. Above the critical temperature the spin (the order parameter) 
expectation value is zero. There are no textures there. Some nonzero density of 
textures can be created during a paramagnet$\rightarrow$ferromagnet transition. 
No textures are created in a quasistatic transition because the correlation 
length diverges at the critical point and the ferromagnetic phase is entered 
already with a long range order (quasi long range order in one spatial 
dimension). However, if the transition preceeds at a finite rate, then, 
according to the Kibble-Zurek scenario, the system passes the critical point out 
of thermal equilibrium. It enters the ferromagnetic phase with a finite 
correlation length. This lack of long range order can manifest itself with 
a finite density of topological textures. We will see that the density of Bloch 
waves created in this way depends not only on the transition rate but also on
the external magnetic field. The leading dependence of the density on the 
transition rate is {\it exponential}, which is much in contrast to the power 
law dependence characteristic for hedgehogs.

\subsection*{Time Dependent Ginzburg-Landau Model}

   Let us define a one dimensional time dependent Ginzburg-Landau model
by the stochastic field equations for the two spin components 
$\phi_{\alpha}(t,x)$, $\alpha=1,2$,

\begin{equation}\label{model}
\dot{\phi}_{\alpha} = 
\phi''_{\alpha} - a(t) \phi_{\alpha} -
[\phi_{\beta}\phi_{\beta}] \phi_{\alpha}
- B\; \delta_{\alpha 2} + \eta_{\alpha}    \;\;,
\end{equation}
where $\dot{}\equiv\partial_{t}$, $ '\equiv\partial_{x}$. 
$\eta_{\alpha}(t,x)$'s are Gaussian noises of temperature $T$ with correlations

\begin{eqnarray}\label{correlations}
&& \langle \eta_{\alpha}(t,x)\rangle=0 \;\;, \nonumber\\
&& \langle \eta_{\alpha}(t_1,x_1)\eta_{\beta}(t_2,x_2)\rangle=
   2 T \delta_{\alpha\beta}\delta(t_1-t_2)\delta(x_1-x_2) \;\;.
\end{eqnarray}
The coefficient $a(t)$ is time dependent. We consider a symmetric linear
quench

\begin{equation}\label{a(t)}
a(t)=
\left\{
\begin{array}{ll}
1                & \mbox{ , if $\; t \leq 0$ }                   \\
1-\frac{t}{\tau} & \mbox{ , if $\; 0 < t \leq 2\tau$ } \\
-1               & \mbox{ , if $\; 2\tau \leq t$ }
\end{array}
\right. 
\end{equation}
Before the quench, for $t<0$, the system is in a symmetric (paramagnetic) 
phase ($a>0$), during the quench, at $t=\tau$, it undergoes a transition
from the symmetric phase ($a(t<\tau)>0$) to a broken symmetry (ferromagnetic)
phase ($a(t>\tau)<0$). Finally it settles down at $a(t)=-1$.

   In the final symmetry broken phase ($a=-1$) the spin field tends
to live at the bottom of the Mexican hat potential 
$V=-|\vec{\phi}|^2/2+|\vec{\phi}|^4/4$. The bottom of the potential is the 
circle $|\phi|=1$. One can consider fluctuations in the direction normal to this 
valley. Their characteristic {\it first relaxation time} is $O(1)$. The 
potential plus the exchange energy generate a length scale $O(1)$ for these 
normal fluctuations. A nonzero magnetic field $B$ removes degeneracy along the 
circle. We assume that $B$ is small so that we can approximately write

\begin{equation}\label{chi}
\phi_1+ i \phi_2= e^{ i \chi + i \frac{\pi}{2} } \;
\end{equation}
for the field at the bottom of the potential valley. 
The spin length is a hard mode so it has a fixed magnitude but its orientation 
(phase $\chi$), which is a soft mode, can vary in space and time. 
Eqs.(\ref{model}) with $a=-1$ and $\eta=0$ reduce to the effective equation

\begin{equation}\label{sineGordon}
\dot{\chi}=\chi''-B\sin\chi \;\;,
\end{equation}
which can be recognized as a diffusive sine-Gordon equation. The length scale 
in this effective model is $B^{-1/2}$ and the {\it second relaxation time} 
is $B^{-1}$. If, as is generic, $B<<1$, then these two scales are much longer 
than those characteristic for the hard magnitude mode. The model 
(\ref{sineGordon}) has a static soliton solution

\begin{equation}\label{kink}
\chi(x)=-4\arctan[\tanh(\frac{B^{1/2}x}{2})]\;\; mod \;\; 2\pi \;\;.
\end{equation}
This soliton is just the Bloch wave.

\subsection*{Instantaneous Quench}

  Once we distinguished between hard and soft modes with corresponding 
length and time scales, we can set to the instantaneous quench ($\tau=0$ in 
Eq.(\ref{a(t)})). In the symmetric phase ($t<\tau$) the order parameter 
fluctuates around the ground state $(\phi_1,\phi_2)=(0,-B)$. If $B<<1$ and the 
temperature $T$ is moderate, then the qubic nonlinearity on the RHS. of 
Eq.(\ref{model}) can be neglected. We are in the Gaussian regime. The system 
before the quench has a unit correlation length. As the quench is instantenous, 
it enters the symmetry broken phase with this correlation length unchanged. Its 
further evolution can be unambigously divided into two stages.

   The first stage is very short, it lasts for $O(1)$ units of time. This is 
the time the order parameter needs to roll down to the bottom of the sombrero 
potential. At the end of this stage the spin already has a fixed magnitude, 
$|\vec{\phi}|\approx 1$. The phase $\chi$ is still chaotic but it is correlated 
over the length scale $O(1)$. The field already contains Bloch waves but they 
are very thin ($O(1)$ width) and distorted, see the plot in Fig.1.

   In the second long stage, which lasts for another $O(B^{-1})$ units of time, 
the initial "baby Bloch waves" grow in size, according to Eq.(\ref{sineGordon}), 
until they become fully pfledged Bloch waves (\ref{kink}), compare Fig.1.

   We want to estimate the density of the Bloch waves after the quench. Let
us first estimate the density but of the baby Bloch waves. We define the
position of the (anti-)baby Bloch wave by a point $x$ where the spin points
up. In the gaussian approximation the field fluctuates around
$(\phi_1,\phi_2)=(0,-B)$ at $t=0^-$. The quadratic potential is turned upside
down at $t=0$. The field at $x$ rolls down 
the slope of the sombrero potential to the upwards spin orientation iff
$\bar{\phi}_2 (x) \; > \; B$ and $\bar{\phi}_1 (x) \;=\; 0$. $\bar{\phi}$ is
an average of $\phi$ over the unit correlation length. In our calculations
below we take this average by introducing an ultraviolet cut-off in momentum
space at $|k|=1$. In the gaussian approximation the probability that the two
conditions hold simultaneously is a product of their probabilities. Let us
work out the probabilities one by one.

  The density of the points such that $\bar{\phi}_{1}(0,x) \;=\; 0$ can
be worked out with the general formula \cite{halperin}

\begin{eqnarray}\label{Ninst}
N[ \bar{\phi}_1 =0]&=&
 \frac{1}{\pi}\sqrt{
 \frac{\langle \bar{\phi}'_{1}(0,x)\bar{\phi}'_{1}(0,x) \rangle}
      {\langle \bar{\phi}_{1}(0,x)\bar{\phi}_{1}(0,x)  \rangle} }=\nonumber\\
&=&\frac{1}{\pi}\sqrt{
 \frac{ \frac{T}{\pi}\int_{-1}^{+1}dk\;\frac{k^2}{1+k^2} }
      { \frac{T}{\pi}\int_{-1}^{+1}dk\;\frac{1}{1+k^2}}  }\approx 0.17 \;\;.
\end{eqnarray}
In the second equality the equilibrium two point correlation function
in momentum space at $t=0^-$ was taken into account.

   Now we estimate the probability that $\bar{\phi}_2(0^-,x)>B$. At $t=0^-$
$\bar{\phi}_{2}$ fluctuates around $-B$. The magnitude of the fluctuation
squared can be found as

\begin{equation}
g^2=\langle [\bar{\phi}_2(0^-,x)+B]^2 \rangle=
    \frac{T}{\pi}\int_{-1}^{+1}dk\; \frac{1}{1+k^2}=\frac{T}{2} \;\;.
\end{equation}
The normalized gaussian probability distribution of $\bar{\phi}_2(0^-,x)$ is
$p(\bar{\phi}_2)=\exp{-(\bar{\phi}_2+B)^2/2g^2}/\sqrt{2\pi g^2}$.
With this distribution the desired probability is

\begin{equation}
P[\bar{\phi}_2>B]\equiv\int_{+B}^{+\infty} p(x)=
\frac{1}{2}\{1-Erf[ \frac{2B}{\sqrt{T}} ]\},
\end{equation}
where we use {\it Mathematica}'s
$Erf[x]=\int_{0}^{x}\frac{2\;dy}{\sqrt{\pi}}\; \exp(-y^2)$. The density
of baby Bloch waves scales like

\begin{equation}
N_{baby}=\; N[ \bar{\phi}_1 =0 ] \times P[\bar{\phi}_2>B] =
         0.08\;c_1\;\{1-Erf[ c_2 \frac{2B}{\sqrt{T}}  ]\}
\end{equation}
with extra parameters $c_1,c_2=O(1)$ to reflect the order-of-magnitude
nature of our calculations.

   $N_{baby}$ does not distinguish between textures and antitextures. Some
pairs of baby textures and antitextures will annihilate. In the second slow
stage of the relaxation process a baby texture grows in width to its full
mature size $\approx 3\times\frac{4}{\sqrt{B}}$. If there is a baby
antitexture within a comparable distance, the texture and antitexture
annihilate each other. An average topological charge density within this
distance is zero. Fluctuation around this average density is the final
density of Bloch waves $N$ and it is proportional to
$\sqrt{ \frac{\sqrt{B}}{12}\times N_{baby} }$

\begin{equation}\label{n}
N=0.08\;C_1\;B^{1/4}
  \sqrt{ 1-Erf[ C_2 \frac{2B}{\sqrt{T}}  ] } \;\;,
\mbox{if $N<<B^{1/2}$}\;\;.
\end{equation}
with constants $C_1,C_2=O(1)$. The formula is accurate in the
dilute regime, $N<<B^{1/2}$. When extrapolated to the oversaturated regime,
it gives an upper estimate. The curve (\ref{n}) favourably compares with the
results of our numerical simulations, see Fig2.

  Density of Bloch waves depends in a critical way on temperature. Very few 
textures are produced for temperatures lower than $ B^2 $. This can
be easily understood. If $T<<B^2$, then just before the quench the field is
localized in a very close neighbourhood of the ground state
$(\phi_1,\phi_2)=(0,-B)$. After the quadratic potential is reversed, the field
uniformly (up to negligible fluctuations) rolls down to $(0,-1)$ without any
chance to wind around the top of the Mexican hat potential.

\subsection*{Finite Rate Quench}

   We first estimate the density of baby Bloch waves at the end of the quench 
at $t=2\tau$. Let us define $m(t)=\langle\phi_{2}(t,x)\rangle$ as a noise 
average of $\phi_2$ or equivalently $\bar{\phi}_2$. A noise average of the
linearized $\alpha=2$ component of Eqs.(\ref{model}) is

\begin{equation}
\dot{m}(t) = - a(t) \; m(t) - B \;\;.
\end{equation}
The solution with the initial condition $m(0)=-B$ and $a(t)$ given by 
(\ref{a(t)}) at the end of the quench is given by

\begin{eqnarray}
m(2\tau) &=& - B - B\sqrt{2\pi\tau} \; e^{\tau/2} \; Erf[\sqrt{\tau/2}] \nonumber\\
         &\stackrel{\tau>>1}{\approx}& - B\sqrt{2\pi\tau} \; e^{\tau/2} \;\;.
\end{eqnarray}
At the end of the quench the fluctuations of $\bar{\phi}_2$ around its
average $m(2\tau)$ are

\begin{eqnarray}
&& g^2 \equiv \langle [\bar{\phi}_{2}(2\tau,x)-m(2\tau)]^2 \rangle =
   \int_{-1}^{+1}dk\; G(2\tau,T,k) \;\;,\nonumber \\
&& G(2\tau,T,k)=\frac{Te^{-4k^2\tau}}{2\pi}
   \{ \frac{1}{1+k^2}+
      \sqrt{\pi\tau}e^{\tau(1+k^2)^2} \times \nonumber\\
&& [ Erf( \sqrt{\tau} (k^2+1) ) - Erf( \sqrt{\tau} (k^2 - 1) ) ]  \}
   \stackrel{\tau>>1,k \neq 1}{\approx}   \nonumber\\
&& \frac{T}{2}\sqrt{\frac{\tau}{\pi}}
   \;e^{\tau(1-2k^2)}\;
   [1-Sign(k^2-1)] \;\;.
\end{eqnarray}
For $\tau>>1$ the fluctuations tend to $g^2\approx T\;e^{\tau}/2\sqrt{2}$.
Similarly as for the instantenous quench the probability that
$\bar{\phi}_2(2\tau,x) \; > \; B$ is given by

\begin{eqnarray}
P[\bar{\phi}_2(2\tau)>B] &=&
\frac{1}{2}\{ 1 - Erf[\frac{B-m(2\tau)}{\sqrt{2\;g^2}}] \} \nonumber\\
&\stackrel{\tau>>\frac{T}{B^2},1}{\approx}&
\sqrt{\frac{ T }{ 8 \sqrt{2} \pi^2 \tau B^2 }}
\exp\{-\frac{2\sqrt{2}\pi \tau B^2 }{T}  \} \;\;.
\end{eqnarray}
At the end of the quench the density of the points such that
$\bar{\phi}_1(2\tau,x)=0$ is given by

\begin{eqnarray}
N[ \bar{\phi}_1 =0]&=&
 \frac{1}{\pi}\sqrt{
 \frac{\langle \bar{\phi}'_{1}(2\tau,x)\bar{\phi}'_{1}(2\tau,x) \rangle}
      {\langle \bar{\phi}_{1}(2\tau,x)\bar{\phi}_{1}(2\tau,x)  \rangle} }=\nonumber\\
&&\frac{1}{\pi}\sqrt{
  \frac{ \int_{-1}^{+1}dk\;k^2\;G(2\tau,T,k) }
       { \int_{-1}^{+1}dk\;G(2\tau,T,k) }  } \stackrel{\tau>>1}{\approx} \nonumber\\
&&\frac{1}{ 2\pi\sqrt{\tau} }\;\;.
\end{eqnarray}
The density of baby Bloch waves is

\begin{equation}
N_{baby}(\tau>>T/B^2)\approx
\sqrt{\frac{ T }{ 32 \sqrt{2} \pi^4 \tau B^2 }}
\exp\{-\frac{2\sqrt{2}\pi \tau B^2 }{T}  \}  \;\;.
\end{equation}
We obtain the density of Bloch waves in a similar way as for the
instantaneous quench,

\begin{eqnarray}\label{N2tau}
N(\tau>>T/B^2)
&=&\sqrt{\frac{\sqrt{B}}{12}\times N_{baby}}= \nonumber \\
&=& 0.035\;D_1\;
    (\frac{T}{B\tau})^{1/4}\;
    \exp\{-D_2\;\frac{\sqrt{2}\pi \tau B^2 }{T}  \} \;\;,
\end{eqnarray}
with constants $D_1,D_2=O(1)$. The density of textures decays
exponentially with $\tau$ on the time scale $O(T/B^2)$. Eq.(\ref{N2tau}) is
compared with numerics in Fig.3.

\subsection*{Conclusion.}

   Almost no textures are produced for $T$ less than $B^2$. Their density decays 
exponentially with the quench time. The time scale for this exponential decay
is of the order of the first relaxation time. 

   Our findings have qualitative implications for higher dimensional textures. 
In the case of Skyrme solitons \cite{skyrme}, which model baryons, an effective 
"magnetic field" is provided by the pion mass term in the Lagrangian. The 
density of baryons just after a chiral transition should be supressed as
compared to the density predicted by the standard Kibble-Zurek 
scenario. This case deserves a separate quantitative study; the study is 
given in a preprint \cite{sadzik}.

   This work extends the Kibble-Zurek scenario to the case with a bias field 
which explicitly breaks the $O(D+1)$ symmetry. In the standard scenario domains 
of characteristic size are formed, each of them has a random orientation chosen 
with uniform probability distribution from the $D$ dimensional sphere. With 
a bias field this probability is not uniform. The landscape just after 
a transition is a uniform sea of spin down with islets of spin up scattered
here and there. With an appropriate winding of spin in its neighbourhood
an islet can survive as a localized topological texture.

\acknowledgements
I would like to thank Leszek Hadasz for help with numerical simulations.
This work was supported by the KBN grant 2 P03B 008 15.

\centerline{\epsfbox{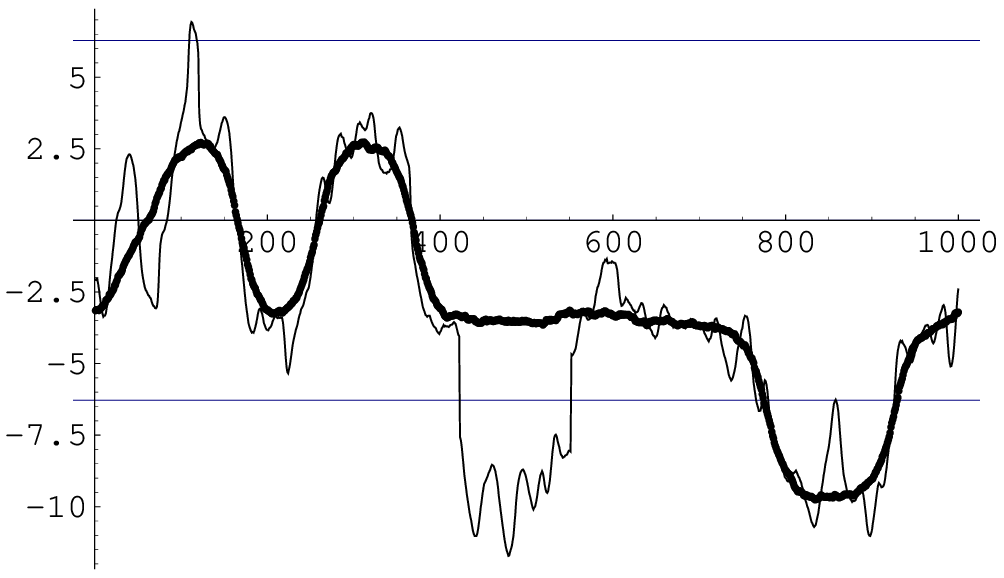}}

{\bf Fig.1. } The phase $\chi$ (compare Eq.(\ref{chi})) as a function
of $x$ for an instantaneous quench with $B=0.005$ and $T=0.00035$.
The thin line is the phase at $t=10$ and the bold line is a plot for $t=1000$. 
A (baby) Bloch wave can be identified as a place where one of the 
$-2\pi,0,+2\pi$ gridlines is crossed. 

\centerline{\epsfbox{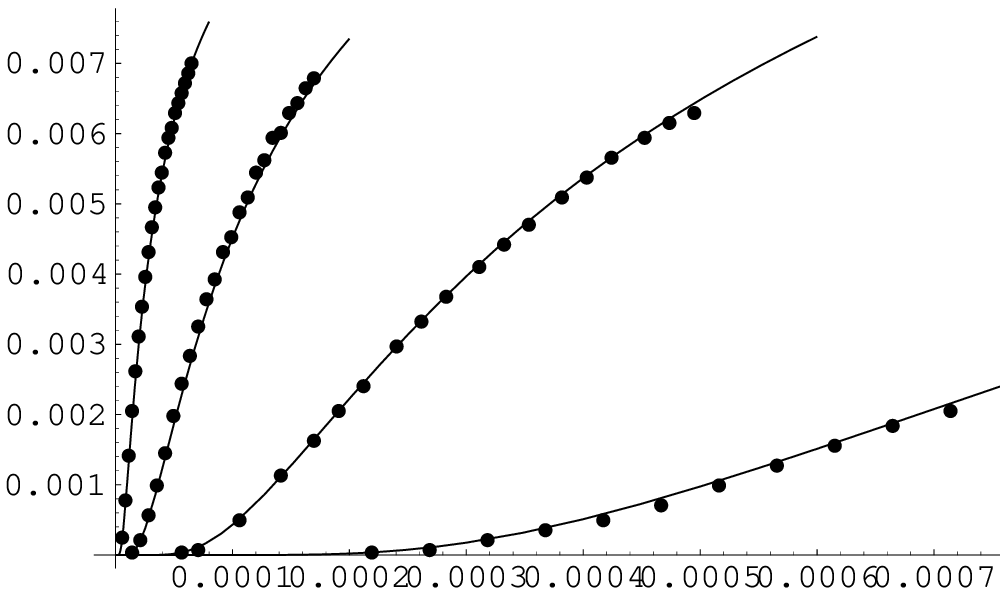}}

{\bf Fig.2. } Average density of Bloch waves as a function of temperature for 
magnetic fields: $0.00125, 0.0025, 0.005, 0.01$ - from left to right.
Each point is an average over several tausend runs with relative error
of $0.3$\% (less than the points size). The lattice constant was $0.2$
and the periodic lattice size was $1000$, which is $30\ldots100$ times
the second correlation length. In each run the solitons were counted
at $t=5/B$ (5 times the second relaxation time), when the soliton
pattern is already stable - compare the bold line in Fig.1.
Numerical data for $B=0.005$ were fitted with the curve (\ref{n})
(solid line) to yield coefficients $C_1=0.88\stackrel{+}{-}0.02$ and 
$C_2=2.45\stackrel{+}{-}0.04$. The errors result from the fit. 
Other solid lines are extrapolation in good agreement with the data.

\centerline{\epsfbox{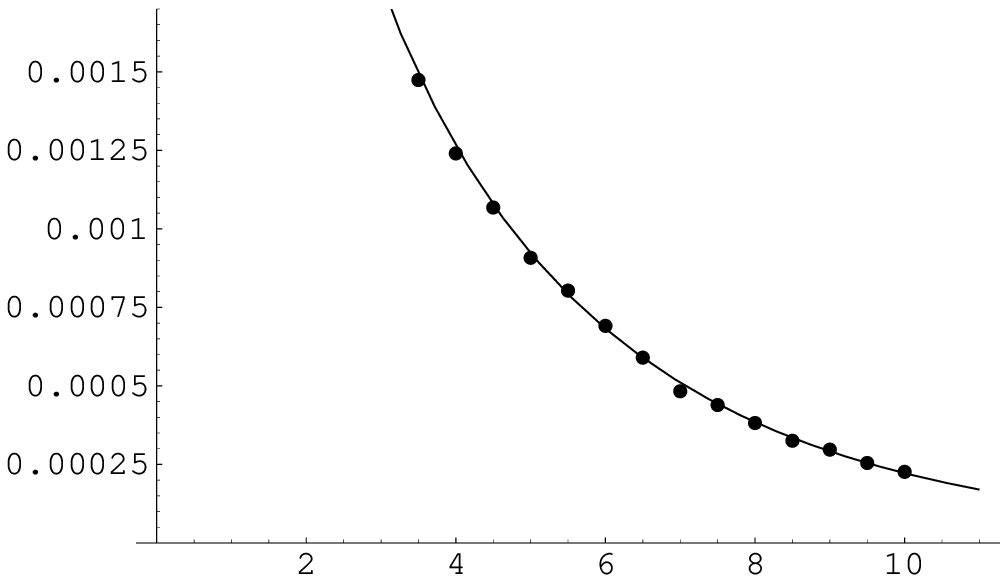}}

{\bf Fig.3. } Density of Bloch waves as a function of the quench time $\tau$ 
for the fixed magnetic field $B=0.005$ and the temperature $T=0.00035$. 
The solitons were counted at $t=2\tau+5/B$. Numerical data were fitted 
with the curve (\ref{N2tau}) - solid line. The coefficients are 
$D_1=0.36\stackrel{+}{-}0.005$ and $D_2=0.74\stackrel{+}{-}0.02$.

\end{document}